# Transitioning to Tomorrow: The Global Journey Towards a Sustainable Energy Economy

Rajini K R Karduri
Department of Civil Engineering
University of Texas, Arlington

*Abstract* - The spotlight is on the intertwined nature of sustainability and energy transition. As the world grapples with environmental challenges, the push for a green approach to energy is more crucial than ever. This transition promises not just a cleaner planet but also better public health and job opportunities. There is a call for united front from policymakers, businesses, and communities to fast-track this eco-friendly shift.

*Keywords—Sustainability; Energy Transition; Ecological Balance; Policymakers*

## I. INTRODUCTION

Sustainability and energy transition are closely linked, addressing the pressing need for a more ecologically conscious approach to energy production and consumption. These initiatives are driven by a combination of social, economic, and environmental factors that collectively guide the journey towards a sustainable energy future. Socially, energy transition offers benefits like improved public health, wider energy access, and community equity. Economically, it provides job opportunities, reduces long-term energy costs, and fosters innovation. Environmentally, the shift to renewable energy leads to reduced carbon emissions, resource conservation, and better air and water quality. Understanding these interconnected elements is crucial for policymakers, businesses, and communities to work together in accelerating the move towards sustainable energy systems, promoting societal well-being, economic growth, and environmental protection.

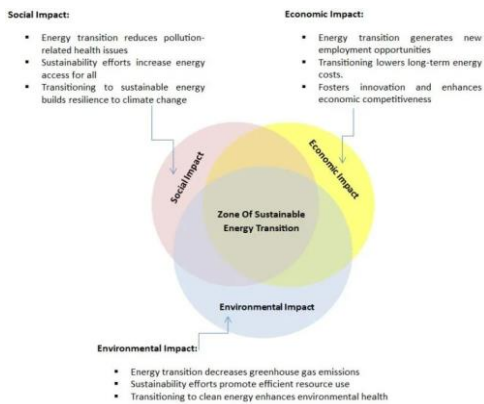

Figure 1. Convergence of social, economic and environmental factors; Credit Source: Author

## II. GLOBAL ECONOMY IMPACTED BY SUSTAINABILITY AND ENERGY TRANSITION

### A. *The Global Shift*

As the world stands at the crossroads of an energy revolution, the global energy landscape is undergoing a profound transformation. Climate change, with its undeniable impacts and looming threats, has thrust the need for sustainable energy into the spotlight. This transition from traditional fossil fuels to renewable energy sources is not just an environmental endeavor; it's a comprehensive economic strategy that holds the potential to redefine global markets. As nations confront the dual challenges of depleting fossil fuel reserves and the environmental repercussions of their consumption, renewable energy emerges as a beacon of hope. This shift promises both ecological balance and economic revitalization. The global community is becoming increasingly aware of the finite nature of fossil fuels and the environmental degradation caused by their excessive use. As a result, there's a growing consensus on the need to find alternative, sustainable energy sources that can meet our needs without compromising the health of our planet. This transition to renewable energy is not just a response to environmental concerns; it's also a strategic move that recognizes the economic potential of green technologies and sustainable practices.

### B. *Economic Opportunities:*

The global pivot towards renewable energy is not just about saving the planet; it's a significant economic powerhouse. Industries centered around renewable sources like solar, wind, and hydropower are burgeoning at an unprecedented rate. These sectors demand a vast, specialized workforce for a multitude of roles, from research and design to installation, operation, and maintenance. This green revolution is creating millions of jobs, revitalizing regions once dependent on traditional industries, and offering a plethora of new career avenues for the global workforce. As these industries grow, they also contribute significantly to the GDP of nations, marking a shift in economic powerhouses. The economic implications of this shift are profound. As countries invest in renewable energy infrastructure, they are also investing in their future economic stability. The renewable energy sector offers a wide range of job opportunities, from technicians and engineers to researchers and policymakers. These jobs not only provide economic security for individuals but also drive economic growth at the national level. Furthermore, as the cost of renewable energy technologies continues to decrease, they present an increasingly viable alternative to traditional fossil fuels, leading to potential savings for both consumers and governments.

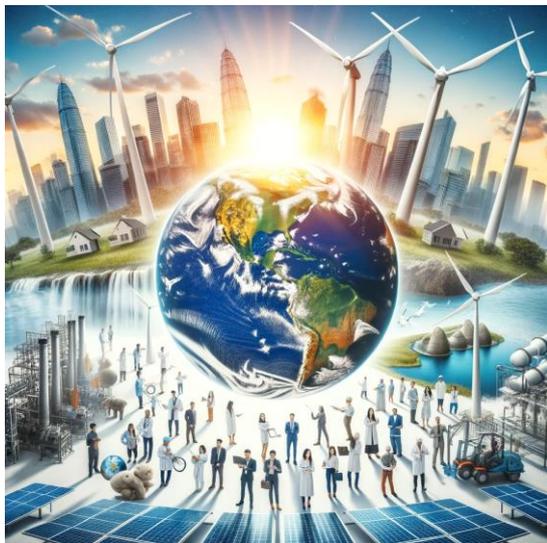

Figure 2: Global shift towards renewable energy, signifying hope and collaboration; Credit Source: Author

*C. Innovation and Research*

The quest for sustainable energy solutions is fueling a renaissance of innovation and research. As nations and corporations race to harness renewable energy, there's a surge in investments in cutting-edge research and development. This drive is leading to breakthroughs in energy storage technologies, smart grid management, and energy efficiency solutions. As these innovations become mainstream, they're not only making renewable energy more accessible but are also setting the stage for technological advancements that could redefine other sectors. The ripple effect of these innovations can be seen across various industries, from transportation to construction, signaling a new era of sustainable development. The energy sector, once dominated by a few major players, is now seeing an influx of startups and entrepreneurs bringing fresh ideas and innovative solutions to the table. This spirit of innovation is propelling the energy sector into a new era, where the focus is not just on meeting energy demands but doing so in the most efficient and sustainable way possible.

*D. Competitive Advantage*

In the global arena, nations that are proactive in their renewable energy endeavors are carving out a niche for themselves. By positioning themselves as leaders in green technologies and sustainable practices, these countries are not only bolstering their domestic economies but are also gaining a formidable edge in international diplomacy and trade. This leadership in sustainability can attract foreign investments, foster strategic trade partnerships, and establish these nations as epicenters of green innovation and thought leadership. As the global market becomes increasingly competitive, nations that prioritize sustainability are likely to emerge as economic leaders in the coming decades. This shift towards green technologies is not just a moral imperative; it's also a strategic one. Countries that invest in renewable energy and sustainable practices are positioning themselves as leaders in a rapidly evolving global market. By doing so, they are attracting foreign investment, fostering innovation, and creating job opportunities, all while reducing their carbon footprint and environmental impact.

*E. Consumer Benefits*

For the everyday consumer, the green energy revolution brings tangible benefits that extend beyond environmental well-being. As the production and distribution of renewable energy become more streamlined and cost-effective, households across the globe can anticipate a significant reduction in their energy bills. This economic relief, coupled with the peace of mind that comes from using eco-friendly energy, promises a future where consumers can enjoy modern amenities without the guilt of environmental degradation. Moreover, as renewable energy becomes more mainstream, consumers will have a wider array of choices, from electric vehicles to solar-powered homes, enhancing their quality of life. The benefits of this shift are manifold. Not only do consumers stand to save money on their energy bills, but they also have the satisfaction of knowing that they are contributing to a more sustainable future. As renewable energy technologies become more accessible and affordable, consumers are also likely to see a wider range of products and services that incorporate these technologies, from smart homes to electric vehicles, further enhancing their quality of life.

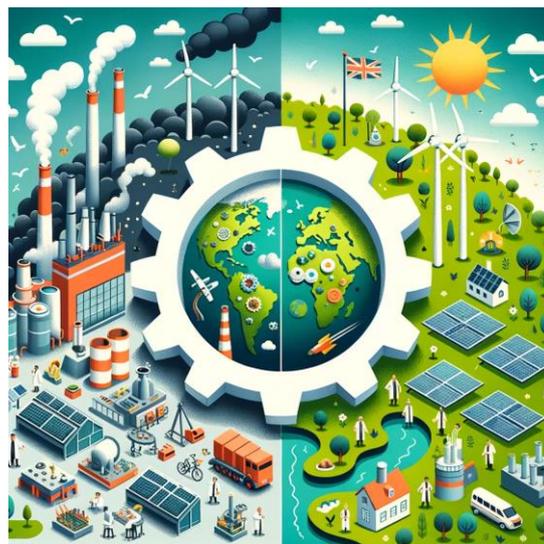

Figure 3: World's transition, contrasting the past with a greener future, emphasizing innovation and international unity;
Credit Source: Author

*F. Challenges Ahead*

The path to a green energy economy, while laden with promise, is not without its hurdles. Existing infrastructure, designed for a world powered by fossil fuels, requires extensive overhauls. Industries steeped in traditional energy practices face the challenges of retraining and adaptation. And while the long-term benefits of green energy are undeniable, the initial capital required for green initiatives can be a significant barrier. Overcoming these challenges demands strategic planning, international collaboration, and unwavering commitment to the vision of a sustainable future. It's a journey that requires patience, resources, and a global perspective. The challenges of transitioning to a green energy economy are manifold. From infrastructural challenges to the need for retraining and reskilling workers, countries face a range of obstacles in their quest for sustainability. However, with the right policies, investments, and international cooperation, these challenges can be overcome, paving the way for a more sustainable and prosperous future.

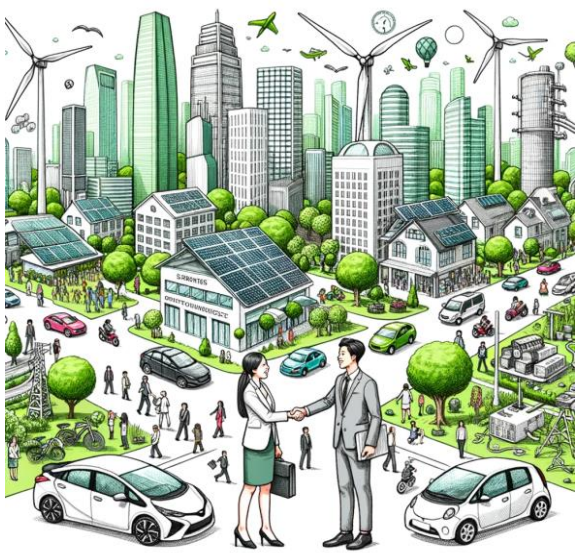

Figure 4: Sustainable urban landscape highlighting the active role businesses play in promoting green practices; Credit Source: Author

### G. Global Collaboration

The energy transition is a global challenge that demands global solutions. International frameworks and agreements, such as the Paris Agreement, underscore the importance of collective action. By uniting under a shared vision, nations can pool resources, share technological advancements, and coordinate policy measures to drive change on a scale that's truly transformative. This spirit of global collaboration is the linchpin for a successful energy transition. As nations come together, sharing successes and learning from failures, the path to a sustainable future becomes clearer and more achievable. The global nature of the energy challenge means that no country can tackle it alone. International cooperation and collaboration are essential if we are to achieve our sustainability goals. By working together, sharing best practices, and pooling resources, countries can accelerate the transition to a green energy economy, benefiting not just their own citizens but the global community as a whole.

### H. Societal Impacts

The societal ramifications of the green energy revolution are profound. Cleaner energy sources promise a drastic reduction in pollution levels, leading to healthier communities. The societal benefits extend to improved public health, reduced medical expenditures, and enhanced overall quality of life. Moreover, as renewable energy becomes more accessible, even remote and underserved communities can hope for a brighter, more connected future. The ripple effects of this transition can be felt in every aspect of society, from education and employment to health and housing, signaling a holistic transformation. The benefits of a green energy revolution extend far beyond the environment. Societies that prioritize sustainability and renewable energy stand to benefit in a myriad of ways. From improved public health outcomes due to reduced pollution to increased economic opportunities in the renewable energy sector, the positive impacts of this transition are far-reaching. Moreover, as renewable energy technologies become more accessible, they have the potential to bridge the energy gap in underserved communities, bringing about positive social change.

### I. The Role of Businesses

In this global narrative, the corporate sector emerges as a pivotal player. Progressive businesses are quickly realizing that green practices are not just about corporate social responsibility; they're a strategic imperative. By integrating sustainability into their core operations and championing green initiatives, businesses can drive change from the ground up. This green corporate movement not only benefits the environment but also resonates with consumers, stakeholders, and investors, offering businesses a competitive edge in the global market. As consumers become more eco-conscious, businesses that prioritize sustainability will likely see increased brand loyalty and profitability. The corporate world has a significant role to play in the global energy transition. From adopting sustainable practices in their operations to investing in renewable energy technologies, businesses can drive change from the ground up. Moreover, as consumers become more environmentally conscious, they are increasingly looking to support businesses that share their values. This shift in consumer behavior presents a significant opportunity for businesses to position themselves as leaders in sustainability, attracting a loyal customer base and driving long-term growth.

### J. The Road Ahead

The journey towards a sustainable energy future is a marathon, not a sprint. It's a complex endeavor that demands patience, perseverance, and vision. But with each step forward, with every innovation, and with every collaborative effort, we move closer to a world where our energy needs are met in harmony with the planet. This transition promises a future that's not just sustainable but is also marked by shared prosperity, innovation, and progress for all. As we look ahead, the challenges may be many, but the rewards - both for our planet and its inhabitants - are immeasurable. The road to a sustainable energy future is long and fraught with challenges. However, with the right policies, investments, and international cooperation, it is a journey that holds the promise of a brighter, more sustainable future. As we navigate the complexities of this transition, it is essential to keep our eyes on the prize - a world where our energy needs are met without compromising the health of our planet or the well-being of its inhabitants.

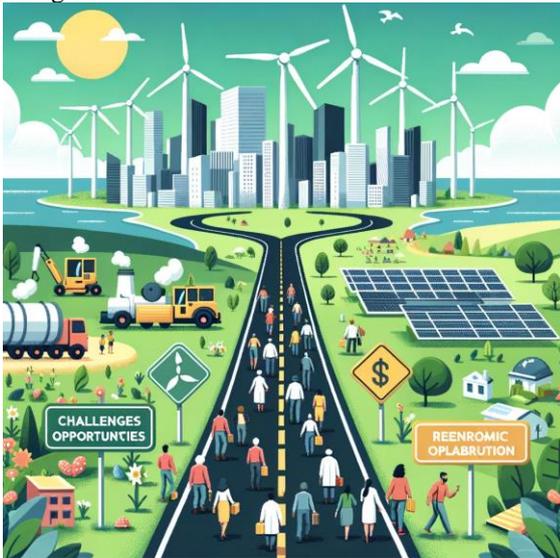

Figure 5: Path towards sustainability, highlighting the challenges and milestones we encounter along the way; Credit Source: Author

### III. SOCIAL IMPLICATIONS OF SUSTAINABILITY AND ENERGY TRANSITION

*A. The Grassroots Movement and Societal Awareness*

The sustainability movement, at its core, began as a grassroots initiative. Concerned citizens, environmental activists, and community leaders recognized the detrimental impacts of unchecked industrial growth and unsustainable practices on the environment. This awareness led to community-driven initiatives, from local recycling programs to urban gardening projects, reflecting a societal shift towards sustainable living. As these local initiatives gained momentum, they sparked a broader societal dialogue about the importance of sustainability, not just as an environmental concern but as a holistic approach to life that encompasses economic, social, and cultural dimensions.

*B. Consumer Behavior and Market Dynamics*

One of the most tangible manifestations of the social influence of sustainability is the change in consumer behavior. Today's consumers are more informed and discerning, often prioritizing products and services that align with their values of sustainability and ethical responsibility. This shift is not merely a trend but a reflection of a deeper societal understanding of the interconnectedness of our actions and their impacts on the environment. Brands and businesses have taken note, with many adopting sustainable practices, not just as a marketing strategy but as a core business principle. This change in market dynamics, driven by consumer demand, is reshaping industries, from fashion to food, emphasizing sustainability, ethical sourcing, and environmental responsibility.

*C. Education and Future Generations*

The discourse around sustainability and energy transition has also permeated educational institutions. Schools and universities worldwide are integrating sustainability into their curricula, recognizing the importance of equipping future generations with the knowledge and skills to navigate the challenges of the 21st century. This educational shift is not limited to academic knowledge; it also encompasses fostering a mindset of responsibility, innovation, and holistic thinking. By instilling these values in young minds, the education system is playing a pivotal role in shaping the future leaders, innovators, and citizens who will champion the cause of sustainability.

*D. Community Cohesion and Shared Responsibility*

The energy transition, particularly the shift towards renewable sources, has also fostered a sense of community cohesion. Localized energy solutions, such as community solar projects or wind farms, not only provide clean energy but also bring communities together in a shared mission. These projects often involve community participation, from planning to implementation, fostering a sense of ownership and shared responsibility. This communal approach to energy solutions is redefining the relationship between individuals, communities, and their environment, emphasizing collaboration, shared benefits, and mutual respect.

*E. Addressing Social Inequalities*

The pursuit of sustainability and energy transition also holds the promise of addressing deep-rooted social inequalities. Access to clean, affordable energy can transform communities, particularly in underserved regions, by powering schools, hospitals, and businesses. Renewable energy projects, with their potential for job creation, can provide economic opportunities in areas previously reliant on declining industries. Moreover, the emphasis on sustainability brings to the forefront issues of environmental justice, recognizing that marginalized communities often bear the brunt of environmental degradation. By addressing these inequalities, the energy transition can pave the way for a more just, equitable, and inclusive society.

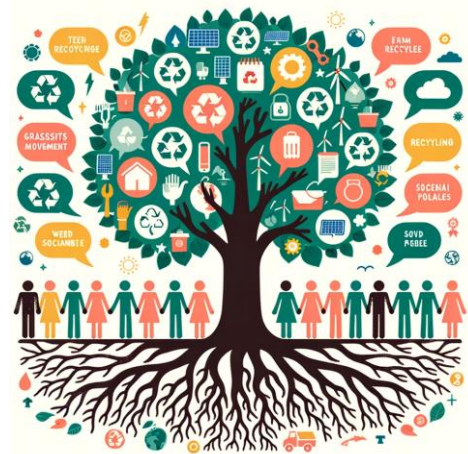

Figure 6: Social sustainability movement;

Credit Source: Author

## IV. ADVANCING ENVIRONMENTAL SUSTAINABILITY AND ENERGY TRANSITION: A COMPREHENSIVE ANALYSIS

The global shift towards environmental sustainability and energy transition is a testament to the collective recognition of the urgent need to address the multifaceted challenges posed by climate change, resource depletion, and environmental degradation. This transition, while rooted in ecological imperatives, has profound implications that span across economic, social, and technological domains.

### A. Renewable Energy and Emission Reductions

Central to the energy transition is the adoption of renewable energy sources, which serve as viable alternatives to fossil fuels. By harnessing the power of nature through solar, wind, and hydropower, we can significantly reduce our carbon footprint. These renewable sources, by their very nature, emit negligible greenhouse gases compared to their fossil fuel counterparts. The shift to renewables, therefore, directly contributes to climate change mitigation, offering a tangible solution to the escalating global carbon emissions crisis.

### B. Resource Efficiency and Ecosystem Preservation

Beyond emission reductions, sustainability initiatives emphasize the judicious use of Earth's resources. The energy transition, by promoting energy efficiency and reducing reliance on non-renewable resources like coal, oil, and natural gas, ensures that we extract less from our planet. This approach not only conserves finite resources but also minimizes the environmental impact associated with their extraction. Such practices protect natural habitats, preserve biodiversity, and prevent the extensive environmental degradation that often accompanies resource extraction.

### C. Enhancing Air and Water Quality

The environmental benefits of the energy transition are not limited to emission reductions and resource conservation. A notable advantage of renewable energy technologies is their minimal contribution to air and water pollution. Unlike the burning of fossil fuels, which releases a plethora of pollutants into the atmosphere, renewables like wind and solar have negligible emissions. This transition promises cleaner air, reduced instances of smog and acid rain, and overall improved air quality. Similarly, sustainable energy practices often incorporate water conservation measures, ensuring that our water sources remain unpolluted and ecosystems thrive.

### D. Localized Impacts and Variability

While the overarching benefits of environmental sustainability and energy transition are universally acknowledged, it's essential to recognize that specific outcomes can vary based on local contexts. The magnitude, nature, and pace of energy transition efforts, coupled with unique environmental and socio-economic factors, can influence the realized benefits in a given region. For instance, an area with abundant sunlight might derive more immediate benefits from solar energy ransition compared to a region with limited sun exposure.

### E. The Road Ahead: Challenges and Opportunities

While the path to environmental sustainability and energy transition is laden with promise, it is not without challenges. From technological hurdles and infrastructural gaps to policy constraints and market dynamics, various factors can impede progress. However, these challenges also present opportunities for innovation, collaboration, and global cooperation. By addressing these roadblocks head-on, leveraging cutting-edge technologies, fostering international partnerships, and galvanizing public and private sector participation, we can accelerate our journey towards a sustainable future.

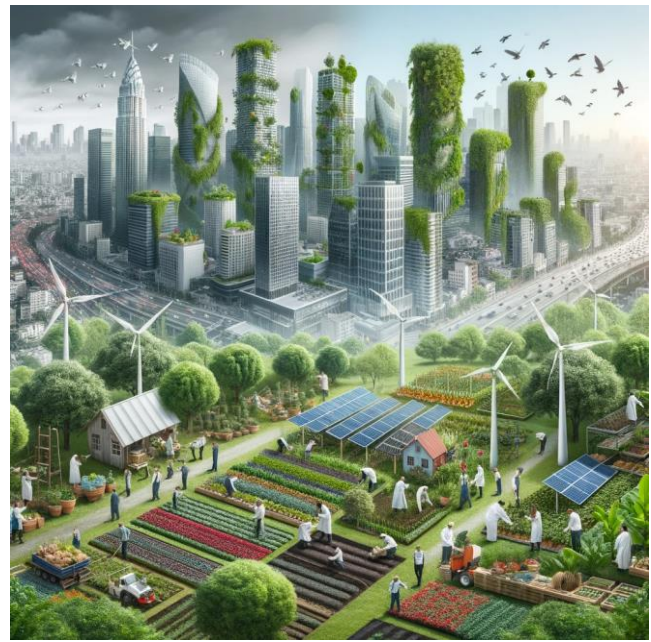

Figure 7: City's evolution from pollution to green initiatives, highlighting the move towards renewable energy and community-driven efforts; Credit Source: Author

## V. CONCLUSION

In an era where the planet's well-being hangs in the balance, the global pivot towards sustainability and energy transition emerges as a beacon of hope. Recent insights from experts in the field reveal a promising tapestry of benefits that this transition brings. From cleaner air in our cities and healthier communities to a surge in green job opportunities and innovations that could reshape our economies, the shift towards renewable energy is more than just an environmental imperative—it's a societal and economic revolution. But like all revolutions, challenges abound. Outdated policies, technological hurdles, and the need for massive public awareness campaigns stand as formidable roadblocks. Yet, with collaboration at the global level and a shared vision for a cleaner, more equitable future, there's every reason to believe that this energy transition can be the catalyst for a brighter tomorrow. As the world watches, the call for action resonates louder than ever.